\documentclass[aps,prx,superscriptaddress,twocolumn]{revtex4-2}
\usepackage{bbm}
\usepackage{mathrsfs}
\usepackage{amsmath}
\usepackage{amsfonts}
\usepackage[colorlinks=true,linkcolor=blue,urlcolor=blue,citecolor=blue,anchorcolor=blue]{hyperref}
\usepackage{graphicx,epstopdf}
\usepackage{subfigure}
\usepackage{epsfig}
\usepackage{dcolumn}
\usepackage{bm}
\usepackage{color}
\usepackage{natbib}
\usepackage{amssymb}
\usepackage{xcolor}
\usepackage{braket}
\usepackage{ulem}
\usepackage{float}
\usepackage{lipsum}

\newcommand{\black}[1]{\textcolor{black}{#1}}
\definecolor{newtxtcolor1}{rgb}{0, 0, 0}
 \newcommand { \ms }[1] {{\color{newtxtcolor1}{#1}}}

\begin{document}

\title{\ms{
Manipulation of Weyl Points in Reciprocal and Nonreciprocal Mechanical Lattices
}}

 \author{Mingsheng Tian}
 \address{State Key Laboratory for Mesoscopic Physics, School of Physics, Frontiers Science Center for Nano-optoelectronics, $\&$ Collaborative Innovation Center of Quantum Matter, Peking University, Beijing 100871, China}
  \address{Department of Physics, University of Illinois at Urbana-Champaign, Urbana, IL 61801-3080, USA}
 \author{Ivan Velkovsky}
   \address{Department of Physics, University of Illinois at Urbana-Champaign, Urbana, IL 61801-3080, USA}
 \author{Tao Chen}
   \address{Department of Physics, University of Illinois at Urbana-Champaign, Urbana, IL 61801-3080, USA}
 \author{Fengxiao Sun}
 \address{State Key Laboratory for Mesoscopic Physics, School of Physics, Frontiers Science Center for Nano-optoelectronics, $\&$ Collaborative Innovation Center of Quantum Matter, Peking University, Beijing 100871, China}

\author{Qiongyi He}	
\email{qiongyihe@pku.edu.cn}
\address{State Key Laboratory for Mesoscopic Physics, School of Physics, Frontiers Science Center for Nano-optoelectronics, $\&$ Collaborative Innovation Center of Quantum Matter, Peking University, Beijing 100871, China}
\address{Collaborative Innovation Center of Extreme Optics, Shanxi University, Taiyuan 030006, China}
\address{Hefei National Laboratory, Hefei, 230088, China}

\author{Bryce Gadway}
\email{bgadway@illinois.edu}
 \address{Department of Physics, University of Illinois at Urbana-Champaign, Urbana, IL 61801-3080, USA}

\begin{abstract}
\ms{
We introduce feedback-measurement technologies to achieve flexible control of Weyl points and conduct the first experimental demonstration of Weyl type I-II transition in mechanical systems. 
We demonstrate that non-Hermiticity can expand the Fermi arc surface states from connecting Weyl points to Weyl rings, and lead to a localization transition of edge states influenced by the interplay between
band topology and the non-Hermitian skin effect.
}
Our findings offer valuable insights into the design and manipulation of Weyl points in mechanical systems, 
providing a promising avenue for manipulating topological modes in non-Hermitian
systems.
\end{abstract}

\maketitle
In recent years, topological phenomena have received considerable attention in photonics and phononics due to their ability to support robust transport against perturbation~\cite{lu2014-npreview,lu2019-rmp,review-phononics,xiujuan-nature2023-review}.
A significant model in three dimensions (3D) is that of Weyl semimetals~\cite{rmp2018weyl,yan2017}, which feature Weyl points (WPs) as linear degeneracies in the band structure.
The topological protection associated with WPs gives rise to intriguing surface states known as ``Fermi arcs"~\cite{naturematerial2016,xu2015-science}, which have been linked to various fascinating phenomena, including chiral anomalies~\cite{nielsen-prb-anomaly,burkov-2015-anomaly}, unconventional superconductivity~\cite{li-prl-superconditivity,xu-2014-np-superconditivity}, and large-volume single-mode lasing~\cite{2012lasing}.
There are two distinct types of WPs: Type-I WPs (WP1) possess a standard cone-like energy spectrum with a point-like Fermi surface, while Type-II WPs (WP2) feature a tilted spectrum with two bands touching at the intersection of electron and hole pockets~\cite{type2-naturereview2015}. Different types of WPs feature distinct transport anisotropies and chiral anomalies~\cite{burkov-2015-anomaly,PhysRevLett.117.086402,PhysRevLett.117.266804,nature2014,weyl2-np2017,weyl2prl2017}.

However, it's often challenging to achieve these distinct features in a specific material due to the difficulty in varying the lattice structure~\cite{weyl-lu2013np,lu2015-science,deng2016-np,xie2019-prl}. 
Toward this goal, synthetic matter is a promising avenue, where the flexible parametric control not only offers versatile means of manipulation but can serve as synthetic dimensions and provide access to
complex lattice designs~\cite{lustig-nature2019-synthetic,ozawa2019-nrp-syn,yuan2018-optica-synthetic,lin2016-nc-2d,qin-2018-oe-2d,wang-prx2017-1d,nguyen-prl2023-synweyl,camelia-prapplied2021,oded-2018nature}.
\ms{
More interestingly, the ability to realize non-Hermiticity provides a method for
exploring diverse non-Hermitian topological properties in Weyl band structure, which has sparked widespread interest~\cite{zhou2018-science-nh,cerjan2019-np-nh,jianchun-prl2022-nonhermitanweyl,litaoprl2023-nonhermitianweyl,rmp2021nh-ep}. 
}

\begin{figure*}[tb]
 \centering
    \includegraphics[width=19cm]{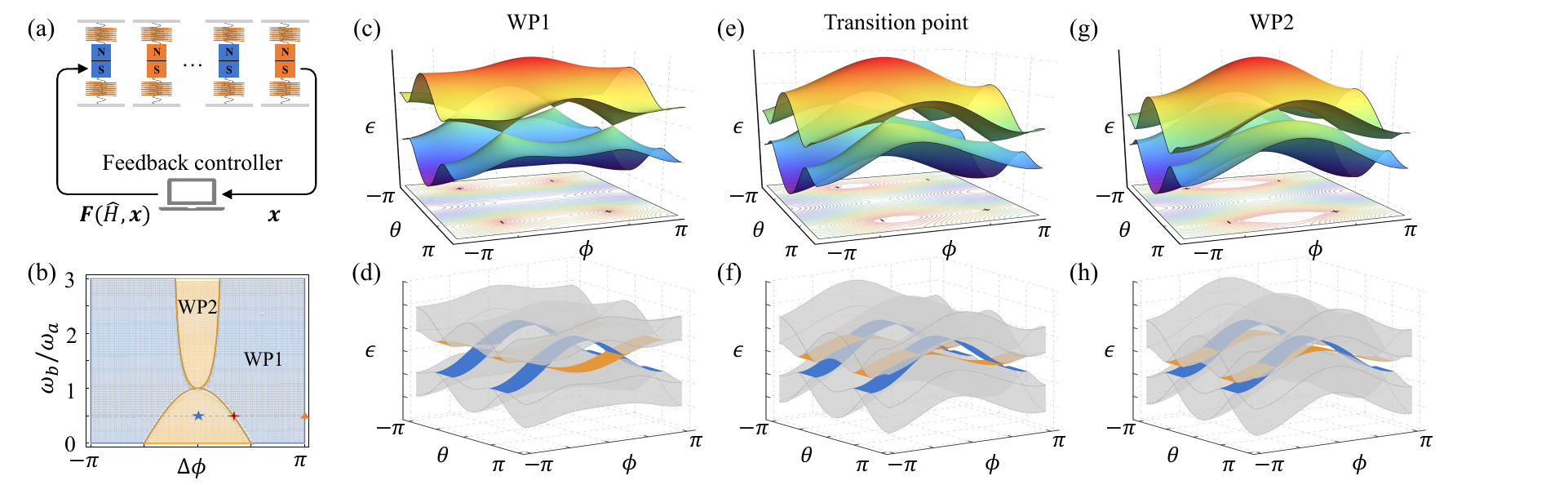}
\caption{Schematic diagram of experimental mechanical lattices and Weyl band structure.
(a)~Illustration of mechanical setup based on measurements and feedback forces.
\ms{
The mechanical arrays consist of N unit cells, each containing sublattice a and sublattice b, marked in blue and orange, respectively.
}
(b)~Phase diagram for WP1 and WP2 as a function of modulation phase difference $\Delta\phi$ and detuning frequency ratio $\omega_b/\omega_a$.
(c)~Band structure of Weyl type-I for periodic mechanical arrays, corresponding to the orange triangle marked in (b) with $\Delta\phi=\pi$ and $\omega_b/\omega_a=1/2$.
The isoenergy contours of the lower band are shown at the bottom of each figure, where we marked the position and chirality of four WPs with signs ``$+$" and ``$-$". 
(d)~The corresponding band structure for truncated mechanical arrays, showing the bulk states (transparent grey)
and Fermi arc surface states (blue and orange colors).
(e,f)~Band structure at Type I-II transition point at the red-cross point of (b).
(g,h)~Type II Weyl band structure at the blue star of (b).
In all plots, we take $\omega_a=1.5 j$, $k=0$, and $\lambda_1=\lambda_2=2j$.
}
\label{fig1}
\end{figure*}

Here, we utilize mechanical oscillators to 
achieve flexible control over WPs.
To generate WPs in three dimensions, we employ measurement-based feedback technologies to construct two additional parametric dimensions in 1D mechanical arrays. 
We experimentally observe, for the first time, the transition between states with WP1 and WP2.
We characterize the transition through: 
i) direct detection of the band structure by Fourier-transformed nonequilibrium dynamics;
ii) direct detection of the Fermi arc surface states under open boundaries.
A unique feature of our mechanical system is its access to non-reciprocal feedback control, which can have significant impacts on the topological transitions through the non-Hermitian skin effect (NHSE).  We show that both the topological phase transition point and the edge localization properties of the topological states are tunable through the non-reciprocal coupling, shedding new light on the interplay between Weyl phase transition and non-Hermitian physics.

\ms{
Our experimental setup includes eighteen mechanical oscillators, whose positions and momenta are
continuously
measured.
We apply individual feedback forces to each oscillator that are responsive to the real-time measurements.
This approach allows us to effectively engineer Hamiltonians of coupled mechanical oscillators by mapping the Newton's equations for feedback-driven oscillators onto a set of
Schrödinger equations of motion~\cite{supplementary}.
Experimentally, we implement feedback forces by using real-time analog output signals to control the currents in gradient coils around each oscillator. This in turn generates magnetic forces on the oscillators, which contain embedded dipole magnets. A desired Hamiltonian, denoted as $h$, is achieved by applying feedback forces
according to Hamilton's equation, $F_i 
 = dp_i/dt = -\partial h / \partial x_i$, where $x_i$ denotes the displacement of the $i$th oscillator.}
Naturally, self-feedback forces proportional to the oscillator positions ($F_i \propto x_i$) shift their frequencies by $\Delta \omega$ from a nominal starting value of $\omega_0/2\pi \approx 13.06~\text{Hz}$. 
Cross-feedback forces related to the nearest oscillators' positions ($F_i \propto x_{i\pm 1}$) introduce independent \ms{and possibly non-reciprocal} left-to-right and right-to-left hopping terms~\cite{supplementary,arxiv-brycesyn,bryce-pre2023}. \ms{The oscillator-specific control of feedback forces also allows for the design of structured, multi-band lattice models.}

\ms{By this approach, we design a Weyl semimetal tight-binding model~\cite{supplementary}, having two-site unit cells with sublattice sites $a$ and $b$,
described in the momentum basis as
\begin{equation}\label{eqhk}
\begin{aligned}
h(\theta,k,\phi)=&
    \begin{pmatrix}
        \omega_0+\omega_a \rm{cos} \phi & 
        \lambda_1\text{cos}\theta+j e^{ik} \\
        \lambda_2\text{cos}\theta+j e^{-ik}  
        &\omega_0+\omega_b \rm{cos} (\phi+\Delta \phi)
    \end{pmatrix} \ .
\end{aligned}
\end{equation}
}
Here, $\omega_0$ represents the on-site natural frequency, while 
$\omega_{a,b}$ are the detuning frequencies, modulated by $\phi$ and $\Delta \phi$.
The off-diagonal terms, $\lambda_{1(2)} \cos\theta$ and $j$,
correspond to intra- and inter-hopping terms, respectively. 
The parameter $k$ signifies to the Bloch wave number in 1D real space.
To construct a 3D Weyl Hamiltonian, we treat $\theta$ and $\phi$ as additional dimensions in the parametric space, combining with $k$ to effectively represent the system as a 3D periodic structure.
In Hermitian systems ($\lambda_1=\lambda_2$), WPs arise by breaking either time-reversal symmetry [$h^*(\mathbf{-q})=h(\mathbf{q})$] or 
parity symmetry [$h(\mathbf{-q})=\sigma_x h(\mathbf{q}) \sigma_x$], 
influenced by the phase difference $\Delta \phi$ and detuning frequency ratio $\omega_a/\omega_b$ of our system, where $\mathbf{q}$ is$(\theta,k,\phi)$~\cite{supplementary}.

Fig.~\ref{fig1}(b) shows the phase diagram for WP1 and WP2 versus the phase difference $\Delta\phi$ and detuning frequency ratio $\omega_b/\omega_a$.
This diagram precisely identifies the transition boundary between WP1 and WP2.
In WP1, a point-like isoenergy surface near WP displays isotropic properties, with isoenergy contours closed and elliptical, as illustrated in Fig.~\ref{fig1}(c).
In contrast, WP2 features `electron' and `hole' pockets touching the isoenergy surface of WP, leading to open, hyperbolic contours~[Fig~\ref{fig1}(g)].
At the transition point [Fig.~\ref{fig1}(e)], the isoenergy contour is a hybrid of hyperbolic and elliptical shapes, indicating unique anisotropic properties.
When truncating the mechanical arrays, two surface states emerge in an open-ended section connecting a pair of WPs, akin to Fermi arcs in electronic systems.
As seen in the band structure of Fig.~\ref{fig1}(d,f,h),
the two strip-like surface states (blue and orange sheets) are well separated from bulk states, intersecting to form four arcs connecting four pairs of WPs.
The clear distinction between the two types of WPs leads to different band dispersions of surface states, allowing us to observe the type I-II transition experimentally.

\begin{figure}[tb]
 \centering
    \includegraphics[width=10cm]{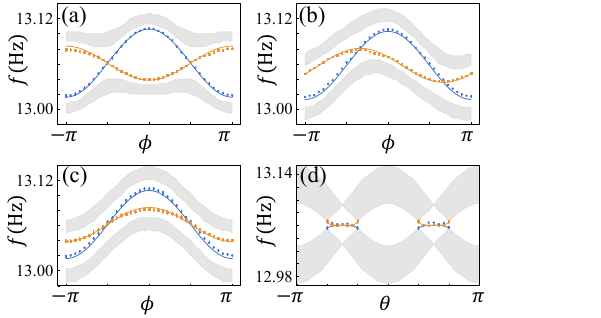}
\caption{Observation of Weyl type I-II transition and Fermi arcs.
(a-c)~Experimentally measured projected band structure in truncated mechanical arrays, showing the case of WP1 (a), the transition point (b), and WP2 (c) at nontrivial cuts of $\theta=0.45\pi$.
(d)~The projected WP1 \ms{($\Delta\phi=\pi$)} band structure dependence on $\theta$ at a special cut of $\phi=\pi/2$ passing through WPs,
where the two edge states appear as arcs connecting a pair of WPs with different chiralities.
Here the orange and blue solid curves represent the two numerical calculated edge states.  
In experiments, we take $\lambda_1/2\pi=\lambda_2/2\pi=60$ mHz,  $j/2\pi=30$ mHz, $\omega_a/2\pi=45$ mHz, $\omega_b/2\pi=22.5$ mHz, and $2N=18$.
The experimental error bars for the edge mode frequencies (typically smaller than the data points) are the standard error based on five measurements.
}
\label{fig2}
\end{figure}

In experiments, we can differentiate WP1 from WP2 by measuring the 
dispersion relations between parameter $\phi$ and frequency $f$,
which is given by the group velocity
$
v_{g}^\phi
=\partial f/\partial \phi
$.
To this aim, we probe the edge modes by beginning with energy only at either the first or last oscillator positions, while we explore the bulk modes by beginning with energy in the ninth site.
By measuring the time evolution of the oscillator dynamics and performing a Fourier transform of the $x_i(t)$ signals, we obtain the relevant frequency spectra as shown in Fig.~\ref{fig2}(a)-(c). 
Here, the orange dots are the peaks of the Fourier spectra of $x_1 (t)$ after initializing at oscillator 1, the blue dots are the peaks of the Fourier spectra of $x_{18} (t)$ after initializing at oscillator 18, and the gray bands are the regions of the Fourier spectra of $x_j (t)$ after initializing at inner oscillator $j$ that have weight above a chosen cutoff value.
For WP1, the group velocities of the two edge states should have opposite directions [Fig.~\ref{fig2}(a)]. 
On the contrary for Type-II WPs, the two group velocities should have the same directions [\ref{fig2}(c)].
And at the transition point, 
one of the edge states manifests a flat dispersion relation with vanishing group velocity, as depicted in Fig.~\ref{fig2}(b).
Additionally, when we measure the projected band structure as a function of $\theta$ along a specific cut that passes through the WPs, 
we observe the Fermi arcs that connect a pair of WPs, as demonstrated in Fig.~\ref{fig2}(d). 
Here, each set of data (points/bands) is acquired from five repeated measurements, incorporating small error bars to depict the minor fluctuations observed in the experiments.
These experimental results align well with the theoretical predictions shown in solid orange and blue curves, exhibiting the first experimental observation of Weyl transition between WP1 and WP2.

\begin{figure}[tb]
 \centering
    \includegraphics[width=10cm]{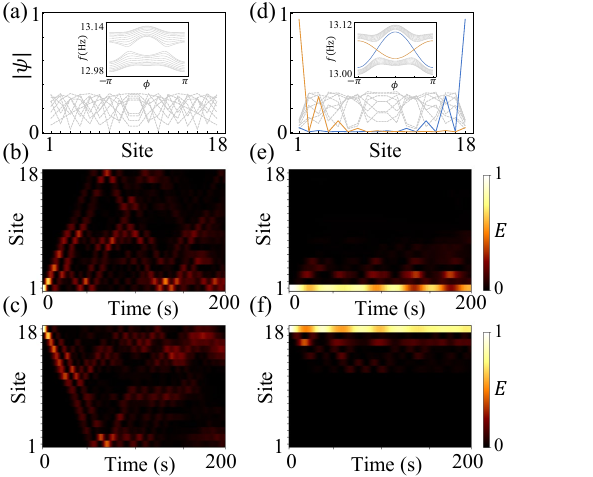}
\caption{Direct observation of the Fermi arc surface states.
(a)~Numerically calculated mode profiles in the trivial region ($\phi=\pi/2$ and $\theta=0.3\pi$).
The inset shows the band structure as a function of $\phi$ when $\theta=0.3\pi$.
(b-c)~Experimentally observed energy distribution of all oscillators evolving with time, where we initially shake either the first (b) or last (d) oscillator. The brightness in these plots reflects the amount of normalized energy at each oscillator. 
(d-f)~Numerically calculated (d) and 
experimentally
observed (e,f) results in the nontrivial region with $\theta=0.45\pi$, showing the presence of edge states.
The orange and blue curves represent the two edge states.
Here, \ms{all figures are plotted in the WP1 region ($\Delta\phi=\pi$)}, and the parameters used in experiments remain consistent with those presented in Fig.~\ref{fig2}.
}
\label{fig3}
\end{figure}

Besides measuring the band structure, the Fermi arc surface states can also be probed through the edge dynamics.
In the experiments, we initialize energy at the first or last
oscillator
and then measure the energy transport throughout the lattice.
If a surface state is present, the oscillators' energies should remain relatively confined to the system’s edge, otherwise, it will extend into the bulk.
In the trivial region
where the parameter $\theta$ is chosen not between two WPs (WPs located at 
$\theta=\pi/3$ and $2\pi/3$),
there is no surface state
[Fig.~\ref{fig3}(a)].
Thus, we find that the initial excitations propagate into the system's bulk as shown in 
Fig.~\ref{fig3}(b,c).
In contrast, in the topologically nontrivial
region
~[Fig.~\ref{fig3}(d)],
the states can be strongly confined to the initial edge.
As shown in
Fig.~\ref{fig3}(e,f),
the states exhibit strong confinement to the edges, resulting in less penetration depths along the truncated directions.

\begin{figure}[tb]
 \centering
    \includegraphics[width=10cm]{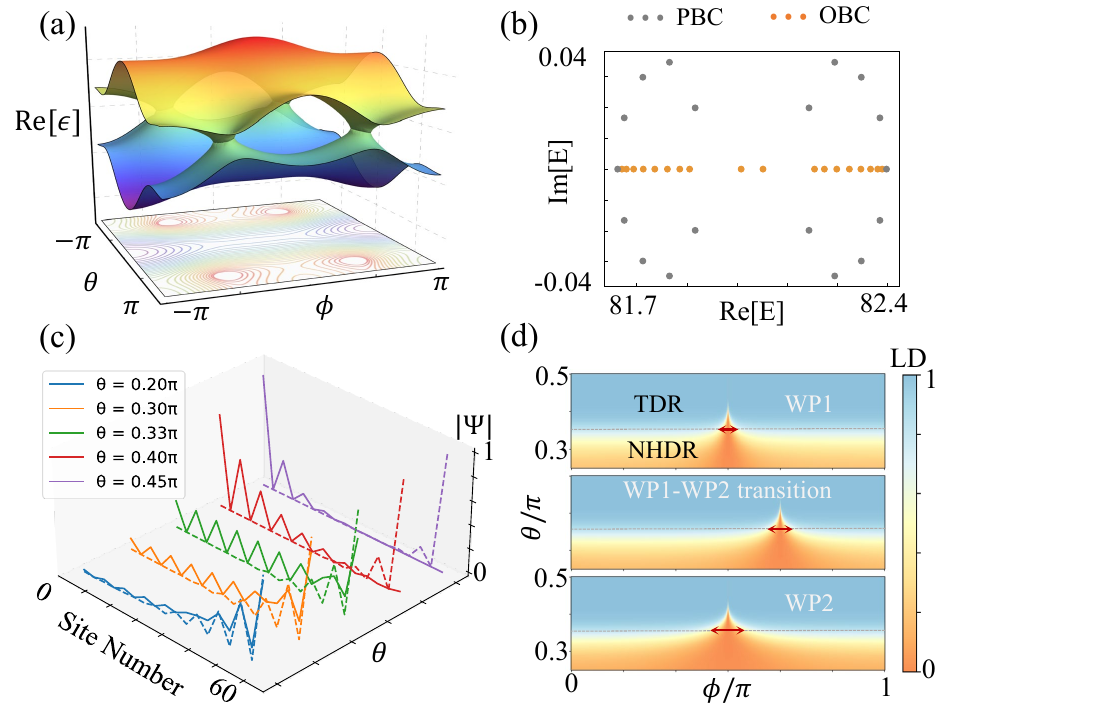}
\caption{
\ms{ The impacts of non-Hermiticity.
(a)~The Weyl ring of non-Hermitian band structure at $\Delta\phi=\pi$.
(b)~The periodic- and open boundary spectra at $\phi=\theta=0.4\pi$.
(c)~The wavefunctions of the edge modes for different $\theta$ at $\phi=0.4\pi$, showing an edge state localization transition from one side to the other. 
(d)~TDR-NHDR diagram of edge state as a function of $\theta$ and $\phi$ for WP1 ($\Delta\phi=\pi$), WP1-WP2 transition point ($\Delta\phi=\pi/3$),  and WP2 ($\Delta\phi=0$). 
In all plots, we take $\omega_a=1.5 j$, $k=0$, $\omega_b/\omega_a=0.5$, and $\lambda_2=2.5 \lambda_1=2j$.
}
}
\label{fig5}
\end{figure}

\begin{figure}[tb]
 \centering
    \includegraphics[width=10cm]{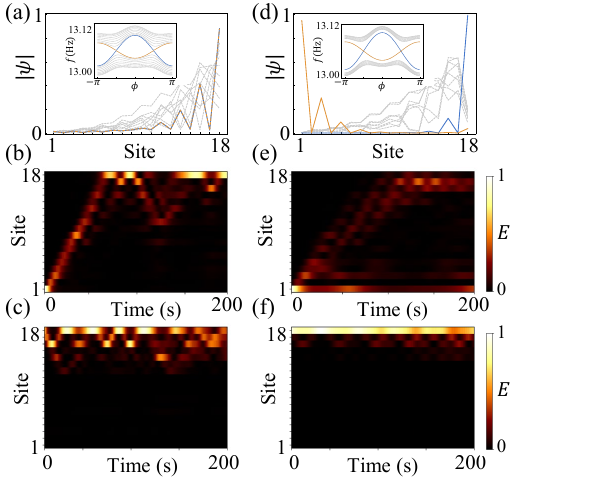}
\caption{Edge localization transition on Fermi arc from the competition of band topology and non-Hermiticity.
We exhibit the numerical mode profiles, band structure, and the experimental time evolution of normalized oscillator energies
with $\theta=0.3\pi$ (a-c) and $\theta=0.45\pi$ (d-f), respectively. 
The surface states exhibit a localization transition, moving from one edge to the opposite edge.
Here, we take
$\lambda_1/2\pi=24$ mHz, $\lambda_2/2\pi=60$ mHz,
and other parameters are same as those depicted in Fig.~\ref{fig3}.
}
\label{fig4}
\end{figure}

\ms{
Going further, by applying a nonreciprocal feedback force between sublattice $a$ and $b$ in each unit cell (\textit{i.e.}, intra-cell hopping $\lambda_1\neq\lambda_2$),
we demonstrate that non-Hermiticity can impact the Weyl band structure in three ways.
First, it transforms Weyl points of band structure into Weyl rings, as shown in Fig.~\ref{fig5}(a) and experimentally explored in the Supplement~\cite{supplementary}. 
Second, the energy spectra under periodic-boundary conditions form loops in
the complex plane, as shown in Fig.~\ref{fig5}(b). This indicates that the NHSE occurs in the bulk modes under open boundary conditions~\cite{zhesenyang-prl2020}. 
Unlike the Hermitian case in Fig.~\ref{fig3}(a), the NHSE modifies the topological transition point, leading to the emergence of surface states appearing at $\theta=0.3\pi$ in Fig.~\ref{fig4}(a).
Third, it changes the localization properties of the Fermi arc surface states.
As shown in Fig.~\ref{fig5}(c), in the area where band topology is dominant, the two edge states localize on opposite sides. In contrast, in regions where skin effects dominate, these states localize on the same edge. 

We define local density of edge states
$
\mathrm{LD}=\sum_{n=1}^{\lfloor N / 2\rfloor}
\sum_{i=a,b}
[|\phi_{n,i}^{e_1}|^2+|\phi_{n,i}^{e_2}|^2]
$
to quantify the topological dominant region (TDR) and the non-Hermitian skin effect dominant region (NHDR). Here, $e_1$ and $e_2$ denote the two edge modes.  
A value of $\mathrm{LD}=1$ indicates that the two states are independently localized at the two chain ends, and an LD value of $0$ implies that the two states are localized at the right chain end.
As depicted in Fig~\ref{fig5}(d),  we plot LD as a function of the parameters $\theta$ and $\phi$, demonstrating the transition from TDR to NHDR for Weyl type I, the Weyl transition point, and type II. 
We note that the peaks, where the skin effects are most pronounced, align with the locations of the Fermi arcs (where two edge states are degenerate). Furthermore, the width of these peaks (depicted in red) increases from WP1 to WP2. 
This is because, at the degenerate point, the
coupling between two edge modes becomes significant, which enhances the skin effect of edge modes and enlarges the non-Hermitian dominant region.
When $\phi$ deviates from the degeneracy point, a gap begins to form between the two edge states, reducing their coupling. Consequently, the area impacted by the skin effect starts to diminish.
Given that WP1 and WP2 possess distinct dispersion relations, the velocities for opening a gap vary significantly. For WP1, the two edge states diverge in opposite directions, whereas for WP2, they move in the same direction. These differences result in a larger gap at WP1 compared to WP2 when $\phi$ deviating from the degeneracy point by the same amount.  A larger gap causes a more pronounced reduction in peak size, explaining why WP1's peak is narrower than that at WP2~(further details in the Supplement~\cite{supplementary}).
}

To observe the edge localization properties in experiments, 
we initially excited energy at the first or last oscillator.
When skin effects are dominant, as seen in Fig.~\ref{fig4}(a)-(c), the
energy initialized
at the first oscillator exhibits a directional flow for short-time dynamics, 
and finally tends to the same edge as the case of initially exciting energy at the last oscillator.
When band topology is dominant (Fig.~\ref{fig4} (d)-(f)),
the energy initialized at the first oscillator in Fig.~\ref{fig4}(e) separates into two distinct parts. 
One tends to flow towards the bulk, displaying a directional flow, and eventually localizes at the upper edge. In contrast, the other part remains localized at the lower edge, together with the excited edge mode seen in Fig.~\ref{fig4}(f), revealing the existence of two different localized mode distributions. 
\ms{
These behaviors differ markedly from those in the Hermitian region. The capability to reshape surface mode wavefunctions enables the engineering of diverse topological modes in the bulk lattice, a development that has recently garnered significant interest~\cite{nature2022morphing,wang2022-prl-morphing}.
}

In conclusion, our study presents a synthetic 3D Weyl model in a mechanical lattice, which was realized by measurement-based feedback.
Through flexible control in parametric space, we successfully observed the transition between WP1 and WP2 and demonstrated the emergence of Fermi-arc states by the observed band structure and edge dynamics in experiments.
Furthermore, by implementing non-reciprocal feedback control, we demonstrate that the non-Hermiticity can impact the topological modes in two ways: i) the Fermi arc surface states are extended from connecting two Weyl points to Weyl rings;
ii) The Fermi arc surface states exhibit localization transition due to the competition between band topology and non-Hermiticity.
These exciting findings not only provide the first experimental demonstration of a Weyl type I-II control but also contribute to the advancement of our understanding of Weyl semimetals in the non-Hermitian region~\cite{yao-prl2018-nh,shanhuifan-science2021,prb2022-competition, xu2017-prl-nhweyl,weidemann2022-nature-topo,xu-pr-2023,prl-dissipation,yogesh-nc2018,yogesh-pra2016,hu-prl2013-nh,adv2020nhphysics}, opening up new avenues for controlling and manipulating topological modes.

\begin{acknowledgments}
We thank Wei Yi at University of Science and Technology of China for the valuable discussions and helpful suggestions. This work is supported by the National Natural Science Foundation of China (Grants No. 12125402, No. 11975026) and the Innovation Program for Quantum Science and Technology (No. 2021ZD0301500) . F. S. acknowledges the China Postdoctoral Science Foundation (Grant No. 2020M680186). I. V., T. C., and B. G. acknowledge support by the National Science Foundation under Grant No. 1945031 and from the AFOSR MURI program under agreement No. FA9550-22-1-0339.
\end{acknowledgments}
% \bibliography{ref}

%

\clearpage
\appendix
\begin{widetext}

%%%%%%%%%%%
\newpage

\section{Experimental implementation}
The theoretical basis for our synthetic mechanical lattices is mapping Newton's equation of motion to the Heisenberg equations for a given Hamiltonian, as described in Ref.~\cite{arxiv-brycesyn}. In this section, we give the theoretical derivation and describe the implementation in our experiments.

The equations of motion for uncoupled and identical harmonic oscillators are:
\begin{equation}
   m \dot{x}_i(t)=p_i(t), \quad \dot{p}_i(t)=-m \omega^2 x_i(t), 
\end{equation}
where $x_i(t)$ and $p_i(t)$ are position and momentum of $i$th oscillator, $\omega$ is the natural oscillation frequency, and $m$ is the mass. 
For simplicity, we introduce the notation $X_i\equiv-\omega^2 x_i(t)$ and $P_i\equiv \frac{-\omega^2}{m}p_i(t)$, so that the equations of motion become
\begin{equation}
   \dot{X}_i=P_i, \quad \dot{P}_i=-\omega^2 X_i. 
\end{equation}
To couple these oscillators as a basis for simulating different Hamiltonians, we now add feedback forces to the system such the equations of motions become
\begin{equation}
    \dot{X}_i=P_i, \quad \dot{P}_i=-\omega^2 X_i+f_i .
\end{equation}
where $f_i$ is a linear function of $(X_1,X_2,\cdots,X_n,P_1,P_2,\cdots,P_n)$.

To map the above equations to Heisenberg equations, we introduce the classical complex variables:
\begin{equation}
\alpha_i \equiv \sqrt{\frac{\omega}{2}} X_i+i \sqrt{\frac{1}{2 \omega}} P_i
\end{equation}
in analogy with the annihilation operator of the quantum harmonic oscillator.
From this, it follows that
\begin{equation}
    X_i=\sqrt{\frac{1}{2 \omega}}\left(\alpha_i+\alpha_i^*\right), \quad P_i=-i \sqrt{\frac{\omega}{2}}\left(\alpha_i-\alpha_i^*\right),
\end{equation}
and hence the equations of motion are
\begin{equation}
    \dot{\alpha}_i=-i \omega \alpha_i+\frac{i}{\sqrt{2 \omega}} f_i
    =-i \omega \alpha_i+\frac{i}{\sqrt{2 \omega}} 
    \sum_j M_{ij}(\alpha_j+\alpha_j^*).
\end{equation}
where the feedback term, $f_i$, is expressed as a linear function of $f_i=\sum_j M_{ij}(\alpha_j+\alpha_j^*)$. To note, this form is specific to the implementation of only $X_j$-dependent feedback forces, relating to real-valued hopping terms and real site-dependent frequency shifts, as utilized in this work.
This equation  can be regarded as the Heisenberg equation of motion derived from a quantum mechanical Hamiltonian (with $\hbar=1$):
\begin{equation}\label{eqs21}
   \hat{H}=\sum_i \omega \hat{\alpha}_i^{\dagger} \hat{\alpha}_i-\frac{1}{\sqrt{2 \omega}} \sum_{i, j}  \hat{\alpha}_i^{\dagger} \hat{M}_{i j} (\hat{\alpha}_j+\hat{\alpha}_j^{\dagger}),
\end{equation}
where $\hat{\alpha}_i^{\dagger}$ and $\hat{\alpha}_i$ are creation and annihilation operators for site $i$ obeying bosonic commutation relations,
with $\alpha_i=\left\langle\hat{\alpha}_i\right\rangle$ and $\alpha_i^*=\left\langle\hat{\alpha}_i^{\dagger}\right\rangle$.

Now, we assume that the feedback is sufficiently weak compared to the natural oscillations. 
Within this limit, the complex amplitudes' time-dependence is still given by $\alpha_i(t) \propto e^{-i \omega t}$.
This allow us to apply the "rotating-wave approximation" (RWA) to simplify Eq.~(\ref{eqs21}):
\begin{equation}\label{eqs21}
   \hat{H}_{\rm RWA}=\sum_i \omega \hat{\alpha}_i^{\dagger} \hat{\alpha}_i-\frac{1}{\sqrt{2 \omega}} \sum_{i, j}  \hat{\alpha}_i^{\dagger} \hat{M}_{i j} \hat{\alpha}_j.
\end{equation}

\begin{figure}[tb]
 \centering
    \includegraphics[width=12cm]{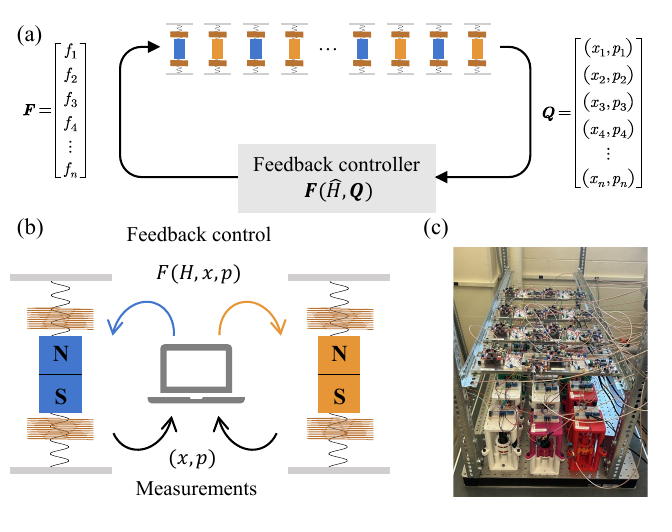}
 \caption{Implementation of our model in a mechanical system via measurement and feedback. 
(a)~Schematic diagrams of our experimental system and approach.
The Hamiltonian can be realized through applied forces that are responsive to real-time measurements. 
\ms{The mechanical arrays consist of N unit cells, each containing sublattice a and sublattice b, marked in blue and orange, respectively.}
(b)~A depiction of the implemented mechanical oscillators, which features embedded accelerometers for the real-time measurement of proxies for the position ($x$) and momentum ($p$), a set of two springs and a dipole magnet embedded in a
pair of anti-Helmholtz coils
for the application of forces. 
Real-time feedback forces $F$, which depend on the real-time measurements $x_i$ and $p_i$, are used to implement the target Hamiltonian $H$. 
(c)~A photograph of our main experimental setup supporting 16 oscillators (2 additional oscillators are located on a separate structure).
}
\label{figs4}
\end{figure}

In experiments, as shown in Fig.~\ref{figs4}, our ``lattice of synthetically coupled oscillators'' consists of modular mechanical oscillators. In the absence of applied feedback forces, these oscillators exhibit nearly identical natural oscillation frequencies,  \black{$\omega_0/2\pi \sim 13.06 \mathrm{~Hz}$.}
Each oscillator is equipped with an analog accelerometer (EVAL-ADXL203). 
Real-time measurements of acceleration $a(t)$ are obtained by sending the signals to a common computer (via very high-gauge wire connections). Additionally, by numerically differentiating the acquired signal, we obtain real-time measurements of the oscillators' jerk, $j(t)\equiv \dot{a}(t)$.
Because $a(t) \propto x(t)$ and $j(t) \propto p(t)$ for a harmonic oscillator, we treat the $a$ and $j$ signals as proxies for the position $x$ and momentum $p$. We additionally multiply (\textit{i.e.}, normalize) the signals to put them on a common scale, reflecting the equipartition of harmonic oscillators.

In this work, $p$-dependent feedback is only used for one purpose, to cancel the natural damping of each of the individual oscillators. By applying self-feedback terms proportional to the oscillator momenta ($p_i$), we cancel the oscillators' natural damping and explore coherent dynamics for well over 500~s ($>6500$ periods). These long timescales are essential for achieving the high frequency resolution of the experimental spectra.

We implement the feedback forces magnetically, avoiding any added mechanical contacts. Each oscillator has a dipole magnet attached to a central, cylindrical shaft. The dipole magnet is embedded in a 
\black{pair of anti-Helmholtz coils.}
We control the current in these coils, which produces an axial magnetic field gradient that in turn creates a force on the oscillator.
We note that we operate the current in a polar fashion, with only one direction of current flow allowed. In the previous iteration of these experiments (Refs.~\cite{arxiv-brycesyn,bryce-pre2023}), we controlled positive and negative variations about an overall offset gradient, thus achieving relative bi-directional control of the forces.
In this updated iteration, we operate with no current offset, instead simply applying feedback currents only when the oscillating force control signals are determined to be positive (\textit{i.e.}, multiplying by an appropriate Heaviside function). The time-averaged effective feedback is the same, as the reduced duty cycle (forces are only applied for a half of each oscillation period) is compensated by an enlarged dynamic range (as no offset current value is applied), while reducing the power consumption of the experiment.

In terms of our data and its analysis, our primary data consists of the real-time measurements of the ${x}$ and ${p}$ signals for each of the oscillators. From these signals, the energy dynamics as presented in Fig.~3 and Fig.~4 of the main text are determined simply by taking the sum of ${x^2}$ and ${p^2}$ (having previously scaled these to signals to be on the same footing).
We can additionally acquire representative energy spectra for the bulk and edge modes of the system by simply taking a Fourier transform \black{(and plotting its absolute value)} of the measured ${x}$ nonequilibrium dynamics signals after applying the appropriate state initialization. As described in the text in relation to Fig.~2, three separate spectra are explored: that of the left edge mode, the right edge mode, and the bulk. The left and right edge modes are explored by initially exciting either the first or last oscillator and then finding the peak of the power spectrum based on the Fourier transform of the $x_1(t)$ or $x_{18}(t)$ signals, respectively. For the bulk modes, where more than just the peak of the spectrum is displayed, the gray band regions of Fig.~2 are obtained as follows. \black{We excite only the ninth oscillator from the system bulk and then perform a Fourier transform of the $x_9(t)$ dynamics. The gray bands plotted in Fig.~2 represent the regions of the resulting power spectra that have weight above a chosen cutoff value. Each of these three spectra is derived from 500~s of nonequilibrium dynamics after the stated initialization.}

\section{Weyl Hamiltonian in synthetic systems}
Before proceeding, let us first provide an introduction to WPs.
WPs in three dimensions exhibit distinct characteristics that set them apart from their two-dimensional counterparts, the Dirac point.
The Dirac cone Hamiltonian in 2D has the form of $h(\mathbf{k})=v_x k_x \sigma_x+v_y k_y \sigma_y$, where $v_i$ are the group velocities.
This form is protected by the product of parity (P) symmetry [$h(\mathbf{k})=\sigma_x h(\mathbf{-k}) \sigma_x$] and time-reversal (T) symmetry [$h^*(\mathbf{-k})=h(\mathbf{k})$].
Any perturbation on $\sigma_z$ terms, which break P or T, will open a band gap.
In comparison, the
Weyl Hamiltonian is expressed as 
$h(\mathbf{k})=v_xk_x\sigma_x+v_y k_y\sigma_y+v_zk_z\sigma_z$.
Since all three Pauli matrices are utilized in the Hamiltonian, it becomes impossible to introduce a perturbation that would open a band gap. This unique characteristic ensures the stability and robustness of WPs in 3D periodic systems.
The elimination and creation of WPs can only occur through pair-annihilations and pair-generations of WPs with opposite chiralities. These processes typically require a strong perturbation in the system. 
The chirality ($c=\pm 1$) of a WP can be defined as $c=\text{sgn}(\text{det}[v_{ij}])$ for
$h(\mathbf{k})=\sum_{ij}k_i v_{ij}\sigma_j$~\cite{weyl-lu2013np}. 

\subsection{Derivation of Weyl Hamiltonian}
\ms{
In our model, the system consists of $N$ unit cells, with each unit cell hosting two sites (sublattice $a$ and $b$), which takes the form of
\begin{equation}\label{seqham3}
\begin{aligned}		
\hat{H}=
&\sum_m^N (\omega_0+ \omega_a \text{cos} \phi ) ~\hat{\alpha}_{m,a}^{\dagger}\hat{\alpha}_{m,a}
+\sum_m^N \left(
\omega_0+ \omega_b\text{cos} (\phi+\Delta\phi)
\right)
~\hat{\alpha}_{m,b}^{\dagger}\hat{\alpha}_{m,b}
\\
&+\sum_m^{N}\left(\lambda_1 \text{cos}\theta ~\hat{\alpha}_{m,a}^{\dagger}\hat{\alpha}_{m,b} +\lambda_2 \text{cos}\theta ~\hat{\alpha}_{m,a} \hat{\alpha}_{m,b}^{\dagger}\right)
+\sum_m^{N-1}\left(j ~\hat{\alpha}_{m+1,a}^{\dagger}\hat{\alpha}_{m,b}+j ~\hat{\alpha}_{m+1,a}\hat{\alpha}_{m,b}^{\dagger}\right),
\end{aligned}	
\end{equation}}where the first two terms represents on-site terms, including natural frequency $\omega_0$, detuning frequency $\omega_{a(b)}, $ and modulating terms $\phi_a$ and $\Delta\phi$.
The middle terms $v_{1(2)}\cos\theta$ represent the intra-cell hopping terms, while the term $j$ corresponds to the inter-hopping terms.
\ms{
Due to the translation invariance of the bulk,  we can make a Fourier transformation $\hat{\alpha}_{k,a(b)}=\frac{1}{\sqrt{N}}\sum_{i=1}^N e^{-i m k} \hat{\alpha}_{m,a(b)}$. 
}
Thus, Eq.~(\ref{seqham3}) can be rewritten in momentum basis
% Under Fourier transformation, Eq.~(\ref{seqham3}) can be rewritten in momentum space
$
H=\sum_k (\hat{\alpha}_{k,a}^{\dagger} \,\,
\hat{\alpha}_{k,b}^{\dagger}) 
~h(k)~
(\hat{\alpha}_{k,a} \,\,
\hat{\alpha}_{{k,b}})^T
$ 
with 
\begin{equation}\label{seqhk}
\begin{aligned}
    h(k) =&
    \begin{pmatrix}
       \omega_0+ \omega_a \text{cos} \phi & \lambda_1\text{cos}\theta+j e^{ik} \\
        \lambda_2 \text{cos}\theta+j e^{-ik} &\omega_0+\omega_b\text{cos} (\phi+\Delta\phi)
    \end{pmatrix}.
\end{aligned}
\end{equation}
where $\sigma_j$, $j=x,y,z$ is Pauli matrix, and the complex parameter $h_j=h_j^R+ih_j^I$ takes the form of 

\begin{equation}\label{eqhk2}
\begin{alignedat}{3}
h_x^R&=\left(\lambda_1 \text{cos}\theta+\lambda_2\text{cos}\theta+2 j\text{cos}k\right)/2,
&\quad&
h_x^I=0,
\\
h_y^R&=-j \text{sin}k,
&\quad&
h_y^I=\left(\lambda_1 \text{cos} \theta-\lambda_2 \text{cos}\theta\right)/2,
\\
h_z^R&=\left(\omega_a \text{cos}\phi_a-\omega_b \text{cos}(\phi+\Delta\phi)\right)/2,
&\quad&
h_z^I=0,
\\
h_0^R&=\left(2 \omega_0+\omega_a \text{cos}\phi_a+\omega_b \text{cos}(\phi+\Delta\phi)\right)/2,
&\quad&
h_0^I=0.
\end{alignedat}
\end{equation}
The complex-energy spectrum of Eq.~(\ref{seqhk}) can be explicitly expressed as follows
\begin{equation}\label{seqeig}
E_{\pm}=h_0\pm \sqrt{ h_R^2-h_I^2+2i ~\mathbf{h_R}\cdot \mathbf{h_I}},
\end{equation}
where $\mathbf{h}=\mathbf{h_R}+i~\mathbf{h_I}$ with $\mathbf{h}\in\mathbb{R}^3$.

For the Hermitian case ($\lambda_1=\lambda_2=v$ with $\mathbf{h_I}=0$), degeneracies in the spectrum [Eq.~\ref{seqeig}] occur only if all
three components of $\mathbf{h_R}$ are simultaneously tuned to zero.
\ms{
Here, $\mathbf{h}$ has one mirror symmetry, $\mathbf{h}(\theta)=\mathbf{h}(-\theta)$, 
and anti-translation symmetry $\mathbf{h}(\phi)=-\mathbf{h}(\phi+\pi)$.
Consider, for instance, a Weyl point at ($\theta$,~k,~$\phi$) with chirality $C=1$. The mirror symmetry reverses the chirality and gives a Weyl point at 
($-\theta$,~k,~$\phi$). The anti-translation symmetry also gives a Weyl point at  ($-\theta$,~k,~$\phi+\pi$) with $C=-1$. Since the Weyl points in our system require $\text{sin}k=0$, with $k=0$ or $\pi$, we conclude that there are a total of 8 Weyl cones in the first Brillouin zone.
}
We assume a WP located at ($\theta_0, k_0, \phi_0$) and 
near the point are 
$\theta_0+q_\theta$, $k_0+q_k$, and $\phi_0+q_\phi$.
In the vicinity of the WP, the Hamiltonian Eq.~(\ref{eqhk2}) should take the Weyl Hamiltonian form~\cite{rmp2018weyl},
and thus can be rewritten as 
\begin{equation}
\begin{alignedat}{1}
h_x&=v \text{cos}\theta_0+j\text{cos}k_0-v \text{sin}\theta_0q_\theta-j \text{sin}k_0q_k
=v_\theta q_\theta,
\\
h_y&=-j \text{sin}k_0+j \text{cos}k_0q_k=v_k q_k,
\\
h_z&=\left(\omega_A \text{cos}\phi_0-\omega_B \text{cos}(\phi_0+\Delta\phi\right)/2
-\omega_A \text{sin}\phi_0q_\phi/2+\omega_B \text{sin}(\phi_0+\Delta\phi)q_\phi/2
=v_\phi q_\phi,
\\
h_0&=\left(2\omega_0+\omega_a \text{cos}\phi_0+\omega_b \text{cos}(\phi_0+\Delta\phi)\right)/2
-\omega_a \text{sin}\phi_0q_\phi/2-\omega_b \text{sin}(\phi_0+\Delta\phi)q_\phi/2
=\epsilon_0+v_0q_\phi,
\end{alignedat}
\end{equation}
where
\begin{equation}\label{seq12}
\begin{alignedat}{3}
v_\theta&=-v \text{sin} \theta_0,
&\quad&
\text{with} \, \text{sin}\theta_0=\pm\frac{\sqrt{v^2-j^2}}{v};
\\
v_k&=j \text{cos} k_0,
&\quad&
\text{with} \, k_0=0,\,\pi;
\\
v_\phi&=-\omega_a \text{sin}\phi_0/2+\omega_b \text{sin}(\phi_0+\Delta\phi)/2,
&\quad&
\text{with} \, \phi_0=\pm \text{arccos}\left(
\mp\frac{\omega_b \text{sin}\Delta\phi}{\sqrt{\omega_a^2+\omega_b^2-2\omega_a\omega_b \text{cos}\Delta\phi}}\right);
\\
v_0&=-\omega_a \text{sin}\phi_0/2-\omega_b \text{sin}(\phi_0+\Delta\phi)/2,
&\quad&
\text{and} \,
\epsilon_0=\left(2\omega_0+\omega_a \text{cos} \phi_0+\omega_b \text{cos} (\phi_0+\Delta\phi)\right)/2.
\end{alignedat}
\end{equation}
The corresponding energy spectrum of the two bands are denoted as follows:
\begin{equation}
\epsilon_{\pm}(\mathbf{q}) = \epsilon_0 + v_0 q_\phi \pm \sqrt{v_\theta^2 q_\theta^2+v_k^2 q_k^2+v_\phi^2 q_\phi^2},
\end{equation}
where ``+'' and ``-'' correspond to the upper and lower bands, respectively. 
In order to distinguish between the two types of WPs, we decompose the Hamiltonian (Eq. (\ref{eqhk2})) into $h_W(\mathbf{q}) = \epsilon_0 \sigma_0 + H_U + H_T$, with
\begin{equation}
\begin{aligned}
h_U&=v_\theta q_\theta \sigma_x+v_k q_k \sigma_y+v_\phi q_\phi \sigma_z, \\
h_T&=v_0 q_\phi  \sigma_0,
\end{aligned}
\end{equation}
where $h_U$ and $h_T$ constitute the potential and kinetic energy components of $h_W(\mathbf{q})$. 
The total energy spectra $\epsilon_{\pm}$ can thus be decomposed into the constant energy component of $\epsilon_0$ as well as the potential and kinetic energy spectra $U$ and $T$, with
\begin{equation}
\begin{aligned}
U=&\sqrt{v_\theta^2 q_\theta^2+v_k^2 q_k^2+v_\phi^2 q_\phi^2}; \\
T=&v_0q_\phi.
\end{aligned}
\end{equation}

\begin{figure*}[t]
 \centering
    \includegraphics[width=17cm]{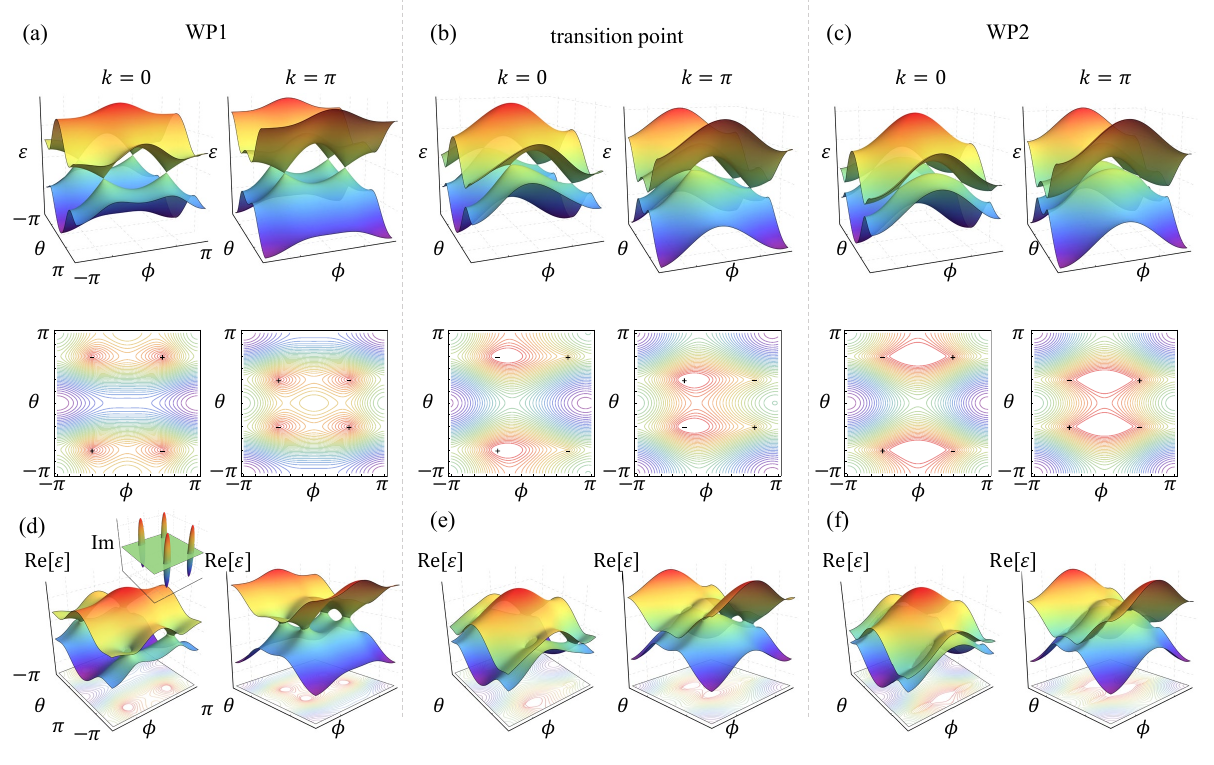}
\caption{(a-c)~Band structure of WP1, transition point, and WP2, where the phase difference satisfies $\Delta\phi=\pi$, $\pi/3$, and $0$, respectively.
The corresponding isoenergy contours of the lower band are shown in the second line, where we marked the position and chirality of four WPs with signs ``+" and ``-".
Here we take $\lambda_1=\lambda_2=2j$, $\omega_a=1.5 j$, and $\omega_b/\omega_a=0.5$.
(d-f)~The real part of the non-Hermitian band structure, with an inset in (d) displaying the band structure of the imaginary part
(the imaginary part band structures in the other figures exhibit the same pattern, hence, for brevity and clarity, we present the results for only one).
Here we take non-reciprocal hopping terms satisfying $\lambda_2=2.5 \lambda_1=2 j$, and other parameters remain consistent with the Hermitian case.
}
\label{figs1}
\end{figure*}

For WP1, $U>T$ should be satisfied in all directions. 
On the contrary, if there exists a particular direction along which $T$ dominates over $U$ with $T>U$, the WPs are of Type-II. Therefore, the system exhibits a phase transition from WP1 to WP2 at $T=U$. The phase transition point can be obtained by comparing $T$ and $U$. 
It can also be obtained by measuring the group velocity in 
$\phi$ parametric space, which is given by 
\begin{equation}
v_{g, \pm}^\phi(q_\theta,q_k,q_\phi)
=\frac{\partial \epsilon_{ \pm}(\mathbf{q})}{\partial q_\phi}
=v_0 \pm \frac{q_\phi v_\phi^2}{\sqrt{v_\theta^2 q_\theta^2+v_k^2q_k^2+v_\phi^2 q_\phi^2}}.
\end{equation}

In the vicinity of WPs with $q_\theta=q_k=0$, the group velocities can be reduced to
$
   v_{g, \pm}^\phi(0,0,q_\phi)=v_0 \pm v_\phi.
$
For WP1, the group velocities of two bands should have opposite directions, which means $\left|v_0\right|<\left|v_\phi\right|$. 
On the contrary for WP2, the two group velocities should have the same direction, such that $\left|v_0\right|>\left|v_\phi\right|$. So at the phase transition point, the group velocity satisfies $\left|v_0\right|=\left|v_\phi\right|$. 
Thus, the two group velocities will become
\begin{equation}
\left\{\begin{array}{l}
v_{\phi,+}=v_0+v_\phi; \\
v_{\phi,-}=0,
\end{array}\right.
\end{equation}
showing that one of the bands will become flat in the vicinity of WPs with zero group velocity while the other band has finite group velocity. So the local flat band structure is a physical signature of phase transition occurring at the boundaries between WP1 and WP2.
In Fig.~\ref{figs1}, we exhibit
the projected band structures and isoenergy contours in
parametric space with k=0 and k=$\pi$. For each type, there are 8 WPs in the first Brillouin zone.

\subsection{Symmetry analysis}
\ms{Weyl points arise from Dirac points by breaking Parity (P) or time-reversal (T) symmetry. 
The Hamiltonian near Dirac point takes the form of $h(\mathbf{k})=v_1 k_1 \sigma_x+v_2 k_2 \sigma_y$, where $v_i$ is the group velocity.
In our model, we introduce the hopping terms among sublattices a and b to realize the Dirac Hamiltonian, which is protected by sublattice symmetry [$h(\mathbf{k})=\sigma_z h(\mathbf{k}) \sigma_z$] and the product of parity symmetry [$h(\mathbf{k})=\sigma_x h(\mathbf{-k}) \sigma_x$] and time-reversal symmetry [$h^*(\mathbf{-k})=h(\mathbf{k})$].
Then we add the phase modulation to realize 
Weyl Hamiltonian in synthetic dimension, expressed as 
$h(\mathbf{k})=v_1 k_1\sigma_x+v_2 k_2\sigma_y+v_3 k_3\sigma_z$.
}

% Time reversal symmetry requires $h^*(\mathbf{-k})=h(\mathbf{k})$, and parity symmetry requires $h(\mathbf{k})=\sigma_x h(\mathbf{-k}) \sigma_x$ in real space with $\mathbf{k}=(k_x,k_y,k_z)$. 
To analyze the Weyl Hamiltonian symmetry in parametric space, , we replace it with $(k,\theta,\phi)$.
In the vicinity of WPs $(k_0,\theta_0,\phi_0)$, we have $(k,\theta,\phi)=(k_0+q_k,\theta_0+q_\theta,\phi_0+q_\phi)$.
For time-reversal symmetry, we have 
\begin{equation}
\begin{aligned}
     h(\theta,k,\phi)&=\sigma_x q_\theta v_\theta+\sigma_y q_k v_k+\sigma_zq_\phi v_\phi
     \\
    h^*(-\theta,-k,-\phi)
    &=\sigma_x q_\theta v_\theta^{\prime}-\sigma_y q_k v_k^{\prime}+\sigma_zq_\phi v_\phi^{\prime}
\end{aligned}
\end{equation}
And  time-reversal symmetry [$h^*(-\theta,-k,-\phi)=h(\theta,k,\phi)$] requires $v_\theta=v_\theta^{\prime}$, $v_k=-v_k^{\prime}$, and $v_\phi=v_\phi^{\prime}$.

For parity symmetry (inversion symmetry), we have
\begin{equation}
    \sigma_x h(-\theta,-k,-\phi)\sigma_x
    =\sigma_x q_\theta v_\theta^{\prime}-\sigma_y q_kv_k^{\prime}-\sigma_zq_\phi v_k^{\prime}
\end{equation}
and parity symmetry [$\sigma_x h(-\theta,-k,-\phi)\sigma_x=h(\theta,k,\phi)$] requires $v_\theta=v_\theta^{\prime}$, $v_k=-v_k^{\prime}$, and $v_\phi=-v_\phi^{\prime}$.

For PT symmetry, we have
\begin{equation}
    \sigma_x h^*(-\theta,-k,-\phi)\sigma_x
    =\sigma_x q_\theta v_\theta^{\prime}+\sigma_y q_k v_k^{\prime}-\sigma_zq_\phi v_\phi^{\prime}
\end{equation}
and PT symmetry [$\sigma_x h^*(-\theta,-k,-\phi)\sigma_x=h(\theta,k,\phi)$] requires $v_\theta=v_\theta^{\prime}$, $v_k=v_k^{\prime}$, and $v_\phi=-v_\phi^{\prime}$.

In our model, as shown in Eq.~(\ref{seq12}), we have $v_\theta=-v \text{sin}\theta_0=
v \text{sin}(-\theta_0)=
v_\theta^{\prime}$, 
$v_k=j \text{cos} k_0
=j \text{cos} (-k_0)=-v_k^{\prime}$, $v_\phi=-\omega_a \text{sin}\phi_0/2+\omega_b \text{sin}(\phi_0+\Delta\phi)/2$, and 
$v_\phi'=\omega_a \text{sin}(-\phi_0)/2-\omega_b \text{sin}(-\phi_0+\Delta\phi)/2$.
For $\Delta\phi \neq 0$ or $\pm \pi$, $v_\phi\neq \pm v_\phi'$, both T- and P-symmetry are broken. 
For $\Delta\phi = 0$ or $\pm \pi$, $v_\phi=v_\phi'$, 
T-symmetry is preserved while P-symmetry is broken,
except for 
$\Delta\phi=0$ and $\omega_a=\omega_b$. In that case, $v_\phi=v_\phi'=0$, and both P- and T-symmetry are protected.
Note that the analysis of T and P symmetry is conducted in the parametric space, and thus can not reflect the true T and P symmetries observed in real space.

\ms{
\subsection{Application of the Weyl type I-II transition}
In Weyl semimetals, type-I and type-II Weyl points exhibit markedly different properties. Specifically, type-II Weyl points are associated with a novel chiral anomaly and \ms{varying densities}, influencing their thermodynamic behaviors distinctively~\cite{type2-naturereview2015}. This is further exemplified in photonic systems, where the diverse band dispersions of these Weyl points enable manipulation of the direction of light propagation. Thus, using flexible parametric control to realize a transition from Weyl type-I to type-II provides a method to manipulate the dynamic behaviors of Weyl semimetals. We would like to present the applications from previous  studies as follows:

     As demonstrated in Ref.~\cite{lin2016-nc-2d}, the band dispersion of edge states can be applied to frequency control and manipulation, such as one-way frequency conversion. In type-I Weyl points, two group velocities ($v_g=\partial E(q)/\partial q$) have opposite signs, while in type-II, both velocities are of the same sign. At the transition point, the emergence of a flat band structure leads to the vanishing of one of the two velocities, thus offering control over the orientation, location, and anisotropy of the Weyl points.

    The isofrequency surfaces for type-I and type-II Weyl points exhibit significant differences. As illustrated in Fig.~\ref{figs5}, these include (a) Type-I with point-like Fermi surfaces; (b) Transition point characterized by unidirectional conical-like Fermi surfaces; and (c) Type-II with symmetric or asymmetric conical-like Fermi surfaces.  Near the Weyl point,
    type-I with point-like surfaces show isotropic properties, whereas type-II displays linear dispersion with a fixed spatial direction $k_1/k_2=\text{const}$. At the transition point, the linear dispersion emerges in only one direction as shown in Fig.~\ref{figs5}(b), revealing unique asymmetric propagation properties.
    These variations in isofrequency surfaces between type-I and type-II result in different light propagation behaviors, as has been theoretically and experimentally
    studied in Ref.~\cite{weyl2prl2017,weyl2-np2017}.

\begin{figure*}[tb]
    \centering
    \includegraphics[width=16cm]{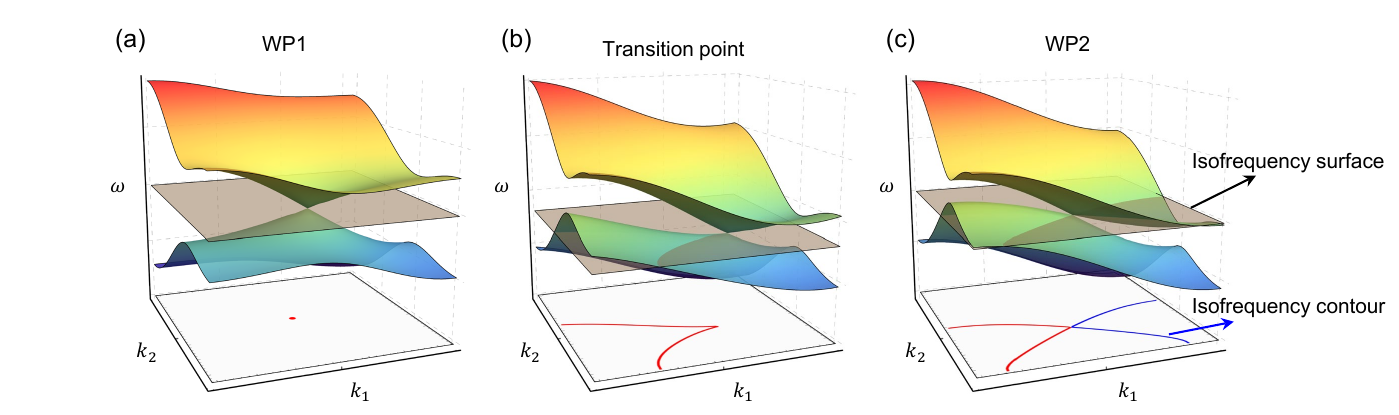}
    \caption{Weyl Dispersion Diagram for Type-I (a), Transition (b), and Type-II (c). The bottom of each figure illustrates the isofrequency contours, with the isofrequency surface passing through the Weyl points. WP1 is characterized by a point-like feature, whereas WP2 displays a crossing at the Weyl point, maintaining a constant spatial direction with $k_1=k_2$=const. At the transition point, linear dispersion emerges solely in one direction, revealing unique asymmetric propagation properties.
    }
    \label{figs5}
\end{figure*}

In our research, we experimentally demonstrate varying dispersion responses of different Weyl types. The flexible Weyl type I-II control gives us a method to modify the dispersion diagram, thus allowing control over the anisotropy of isofrequency contours around Weyl points.
This paves the way for further applications in manipulating dynamic behaviors in Weyl semimetals.  

}

\section{Impact of non-Hermiticity in Weyl Hamiltonian}\label{Weylring}

\subsection{Non-Hermitian band structure}
\ms{
The non-reciprocal hopping terms are introduced in our system by nonreciprocal feedback forces. Specifically, a desired Hamiltonian, denoted as $h$, is achieved by applying feedback forces to the oscillators in the form $F_i=dp_i/dt=-\partial h / \partial x_i$, where $x_i$ denotes the displacement of the $i$th oscillator.
Cross-feedback forces related to the nearest oscillators' positions allow us to introduce independent and possibly non-reciprocal left-to-right ($F_i \propto x_{i+1}$) and right-to-left ($F_{i+1}\propto x_{i}$) hopping terms.
In our model, the non-Hermiticity is achieved by applying a nonreciprocal feedback force between sublattice sites a and b in each unit cell, with intra-cell hopping term $\lambda_1\neq\lambda_2$.
With breaking parity symmetry, degeneracies occur when $\mathbf{h}_R^2-\mathbf{h}_I^2=0$ and $\mathbf{h}_R\cdot \mathbf{h}_I=0$ are simultaneously satisfied [see Eq.~(\ref{seqeig})]. These conditions transform the Weyl points into Weyl rings in the $\theta$-$\phi$ space.
}

As shown in Fig.~\ref{figs2}(a,b), we plot
the real and imaginary band structure when $\lambda_1\neq \lambda_2$ (due to the mirror symmetry, it suffices to showcase the portion of parameters ranging from 0 to $\pi$).
It is observable that the degeneracy points form a nodal line in the parametric space, indicating a transformation of the WPs from the Hermitian band structure into a Weyl ring in the non-Hermitian system. The presence of this Weyl ring in the non-Hermitian region can be experimentally confirmed within our systems.
As depicted in Fig.~\ref{figs2}(c,d), we examine the projected band structure using a $\theta$ cut and $\phi$ cut. 

\black{For the real parts we initialize on the left site, measure the dynamics for a total of 500~s, and find the peaks of the fourier spectra of $x_1(t)$. For the imaginary parts, we fit the $x_1^2 + p_1^2$ dynamical growth to the form $e^{4\pi g t}$ and relate
}
\black{$g = \textrm{Im}[\varepsilon]$}
(note that the negative imaginary parts, which indicate the decay rate of oscillators, are not shown in our plots as the competition between gain and loss makes it difficult to measure the decay rate). 
In the PT-phase symmetric region, it is noticeable that the imaginary part of the system is zero. However, in the PT-phase broken region, the imaginary part becomes non-zero, leading to a significant increase or decay in the energy of the oscillator.
The bifurcation structure observed in both $\theta$ and $\phi$ closely align with the theoretical prediction (black and gray curves). This correspondence suggests that the observed bifurcation patterns serve as evidence for the existence of Weyl rings in the non-Hermitian regions.

\begin{figure*}[t]
 \centering
    \includegraphics[width=15cm]{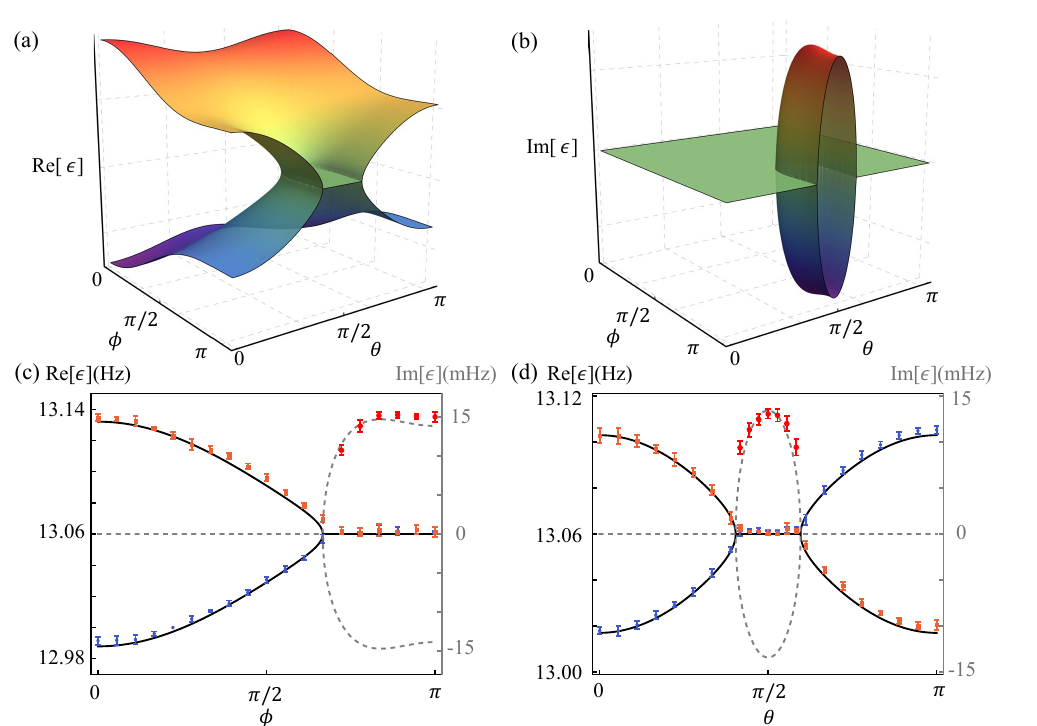}
\caption{Non-Hermitian band structure. (a,b)~The real and imaginary parts of the band structure are plotted in parametric space.
The experimental results are shown through a $\theta=\pi/2$ cut in (c) and a $\phi=\pi$ cut in (d), where the orange and blue dots are real parts and the red dots represent the positive imaginary parts.
The black and gray curves are the corresponding theoretical results. 
In experiments, we set $k=0$, $\Delta \phi=\pi$, $\lambda_1/2\pi=24$~mHz, $\lambda_2/2\pi=60$~mHz, $j/2\pi=30$~mHz, and $\omega_a/2\pi=\omega_b/2\pi=45$~mHz.
Each data point here is acquired from five repeated measurements in two oscillators, with the error bars representing the standard error of those sets of measurements.}
\label{figs2}
\end{figure*}

\ms{
\subsection{Non-Hermitian skin effects and edge state localization transition}
\begin{figure*}[tb]
    \centering
    \includegraphics[width=16cm]{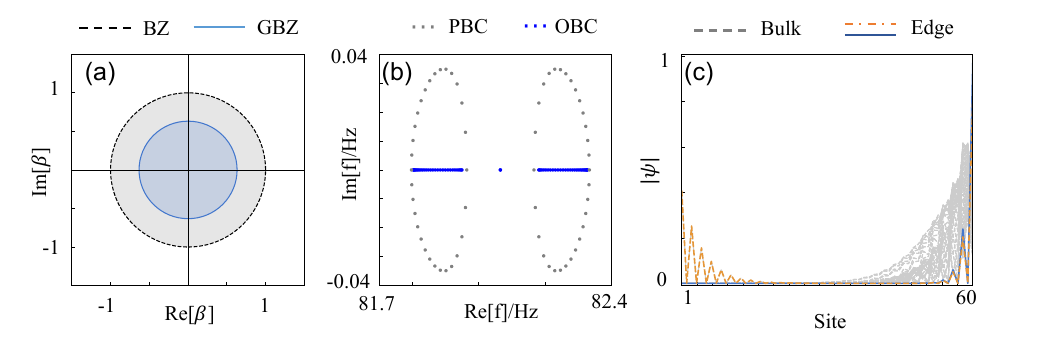}
    \caption{BZ and GBZ (a), periodic- and open-boundary spectra
    (b), and all normalized eigenfunctions for open boundary (c) are plotted.
    Here, we take $\Delta \phi=\pi$, $\theta=0.4\pi$, $\phi=0.5\pi $, $\lambda_1/2\pi=24$~mHz, $\lambda_2/2\pi=60$~mHz, $j/2\pi=30$~mHz, and $\omega_a/2\pi=\omega_b/2\pi=45$~mHz. 
    While we emphasize results in the WP1 region, these phenomena also occur in WP2 and WP1-WP2 transition points.
    }
    \label{figs6}
\end{figure*}
One of the intriguing and distinct features of non-Hermitian systems is the presence of the non-Hermitian skin effect when truncating to open boundary conditions, which will lead all modes to localize on the same edge.
In non-Hermitian systems, the sufficient and necessary conditions for the skin effect are~\cite{zhesenyang-prl2020}: (i) Any part of the generalized Brillouin zone (GBZ) that lies within (without) the unit circle corresponds to a set of skin modes on the right (left) side. (ii) The skin effects occur if and only if the periodic spectra enclose a finite area. In our cases, the nonreciprocal hopping terms break parity symmetry,
resulting in the entire GBZ being inside the unit circle [Fig.~\ref{figs6}(a)] and the periodic spectra's shape enclosing a finite area [Fig.~\ref{figs6} (b)]. Consequently, all bulk states manifest as skin modes on the right side, as depicted in Fig.~\ref{figs6}(c).

We demonstrate that the non-Hermitian skin effect (NHSE) can impact the Weyl band structure in two ways.
First, it modifies the topological transition point (see Fig.~3(a) and Fig.~5(a) of main text). 
Second, it changes the edge localization of the Fermi arc surface states.
With the skin effect, the bulk modes tend to be
localized on the right side. In contrast, the edge states show a different localization due to the competition between band symmetry and skin effects.
In the topologically nontrivial area where symmetry is dominant, the two edge states localize on opposite sides. In contrast, when the skin effect is dominant, these states localize on the same edge. The interplay between these two dynamics leads to a localization transition of the edge states.

To quantify the topological dominant region (TDR) and the non-Hermitian skin effect dominant region (NHDR), we use a local density (LD) of edge modes over the left-half cells, defined as:
\begin{equation}
\label{seqLD}
\mathrm{LD}=\sum_{n=1}^{\lfloor N / 2\rfloor}
\sum_{i=a,b}
[|\phi_{n,i}^{e_1}|^2+|\phi_{n,i}^{e_2}|^2],
\end{equation}
where $\lfloor N / 2\rfloor$ denotes the floor function rounding down $N / 2$ to its next lowest integer. The superscripts $e_1$ and $e_2$ refer to the two edge modes. 
The quantity LD serves as a count for the number of localized states at the left end of the chain.
Specifically, a value of $\mathrm{LD}=1$ indicates that the two states are independently localized at the two chain ends, signifying the occurrence of the TDR. On the other hand, an LD value of $0$ implies that the two states are localized at the right chain end, representing the presence of edge modes associated with the NHDR.

\begin{figure*}[tb]
    \centering
    \includegraphics[width=16cm]{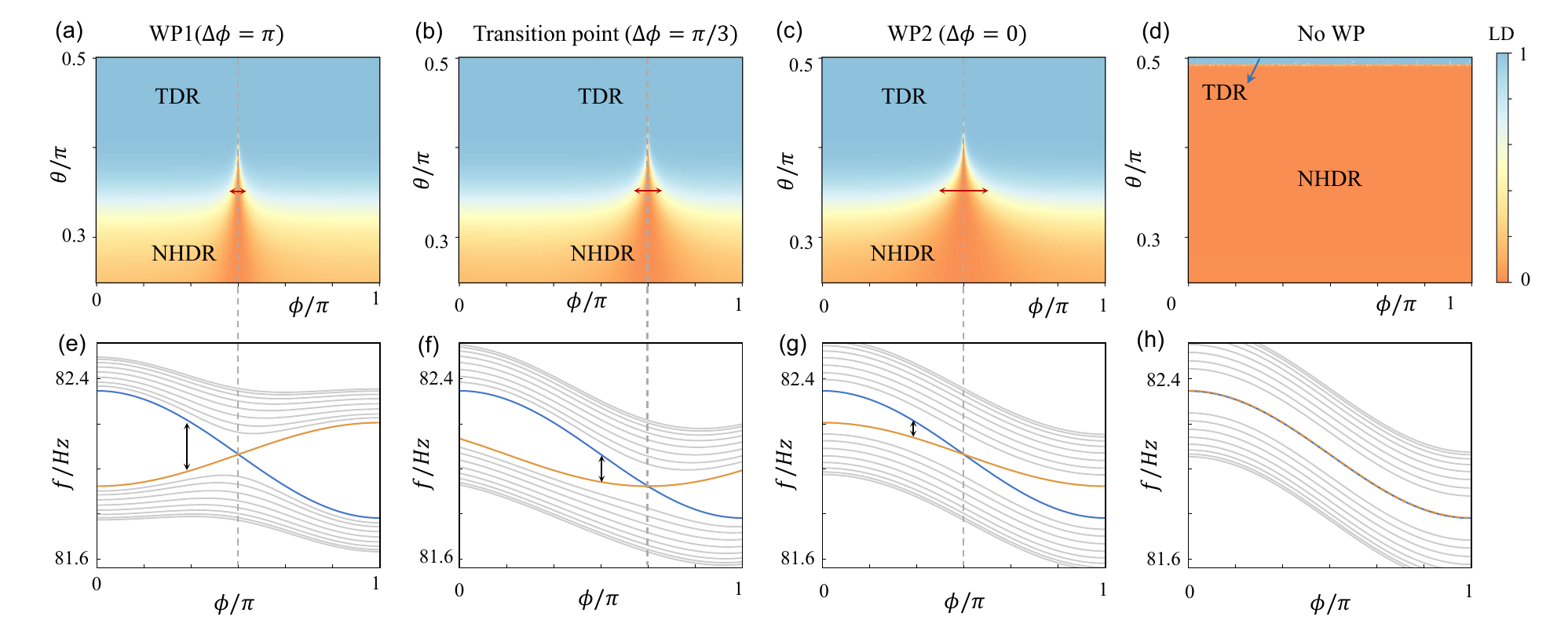}
    \caption{Influence of skin effects on WP1 and WP2.
    (a)-(c) TDR-NHDR diagram of edge state localization transition as a function of system size $\phi$ and $\theta$ for WP1 (a), Weyl transition point (b), and WP2 (c).
TDR (blue region) and NHDR (orange region) are characterized by $\rm{LD}=0$, and 1, respectively.
The peaks marked by dashed grey lines are the location of the Fermi arc, where the eigenvalues of two edge states are degenerate.
(e)-(g) The corresponding energy spectrum at $\theta=0.35\pi$.
Here we take $\lambda_2=2 j=2.5 \lambda_1$, $\omega_a=2 \omega_b=1.5 j$, and $2N=18$.
(d)(h) A specific scenario with $\omega_a =\omega_b $ and $\Delta\phi = 0$. The NHDR is significantly larger than the TDR, and the two edge states are degenerate for any value of $\phi$.
    }
    \label{figs7}
\end{figure*}

\begin{figure*}[t]
 \centering
    \includegraphics[width=17cm]{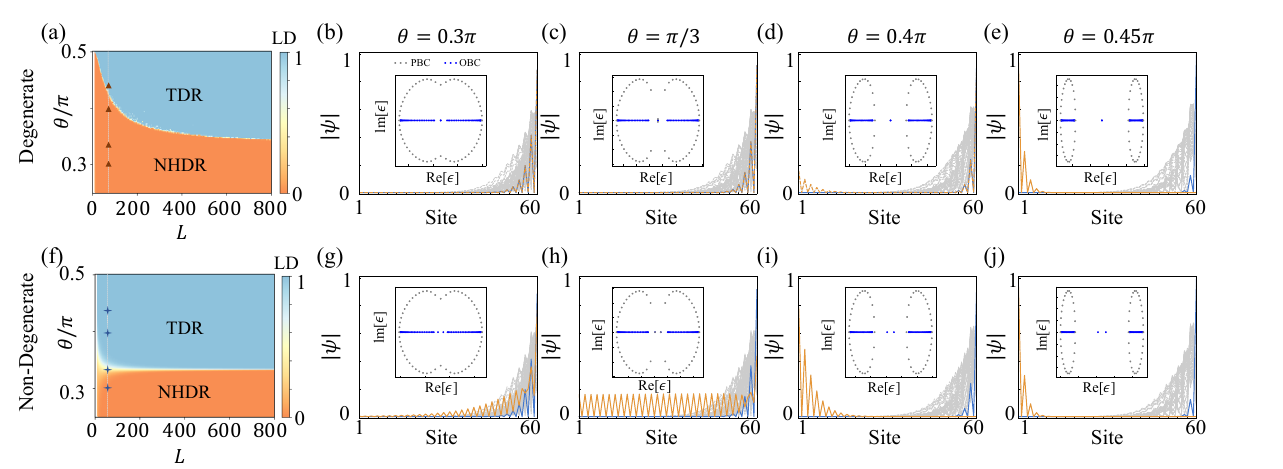}
\caption{The edge state transition in finite-size systems. 
(a)~ Numerically calculated TDR-NHDR diagram of edge state transition as a function of system size $L$ and
$\theta$ when two edge states are degenerate.
TDR (blue region) and NHDR (orange region) are characterized by $\rm{LD}=0$, and 1, respectively.
The dashed line with the red triangles denotes four examples at $L=60$ as illustrated in (b)-(e), where the mode profiles for two surface modes (blue and orange) and bulk modes (gray) are presented.
The insets show the corresponding PBC and OBC spectra of the systems.
Here we take $\lambda_2=2 j=2.5 \lambda_1$, $\omega_a=2 \omega_b=1.5 j$, $\phi=\pi/2$,  and $\Delta\phi=0$,
(f)-(j)~Similar studies are plotted at $\phi=0.4 \pi$ with two edge states being non-degenerate,
revealing the robustness of edge state to system size and the emergence of delocalized surface modes at transition points (h).
Here we take $\phi=0.4 \pi$, and other parameters are the same as those presented in (a)-(e).
}
\label{figs3}
\end{figure*}

The transition points and localization characteristics differ between type-I and type-II Weyl points. 
As depicted in Fig.~\ref{figs7}, we plot the LD as a function of the parameters $\theta$ and $\phi$, demonstrating the transition from topological dominated region (TDR) to non-Hermitian skin effect dominated region (NHDR) for Weyl type I (a), the Weyl transition point (b), and type II (c). 
We note that the peaks, where the skin effects are most pronounced (dashed grey lines), align with the locations of the Fermi arcs (where two edge states are degenerate). Furthermore, the width of these peaks (depicted in red) increases from WP1 to WP2. This widening is due to the varying band dispersion. A detailed explanation is provided below:

At the degenerate point, the coupling between two edge modes becomes significant.
This coupling enhances the skin effect of edge modes and enlarges the non-Hermitian dominant region.
When $\phi$ deviates from the degeneracy point, a gap begins to form between the two edge states, reducing their coupling. Consequently, the area impacted by the skin effect starts to diminish.
Given that WP1 and WP2 possess distinct dispersion relations, the velocities for opening a gap vary significantly. For WP1, the two edge states deviate in opposite directions, whereas for WP2, they move in the same direction. These differences result in a larger gap at WP1 compared to WP2 when deviating from the degeneracy point by the same amount, as illustrated by the black bidirectional arrow in Fig.~\ref{figs7} (e)-(g). A larger gap causes a more pronounced reduction in peak size, explaining why WP1's peak is narrower than that at WP2. Interestingly, in a particular scenario where two edge states are degenerate for any value of $\phi$ (Fig.~\ref{figs7}(h)), most of the region is dominated by skin effects, as illustrated in Fig.~\ref{figs7}(d).

To further understand these behaviors, we provide the exact calculation of edge states as detailed in Sec.~\ref{exactEdge} of the supplement.
Our findings reveal that when the eigenvalues of two edge states are degenerate, the transition point occurs at $\theta=\pi/3$ in the thermodynamic limit. In this scenario, the two edge states are separately located at opposite ends of the chain, with no overlap or coupling.  However, in systems with a finite number of cells, the coupling between edge states becomes significant. This coupling enhances the skin effect of the edge modes and enlarges the non-Hermitian dominant region. 
As a result, the NHDR expands when the system size is reduced, as shown in Figure~\ref{figs3}(a).
In contrast, when the two edge states are non-degenerate, as demonstrated in Fig.~\ref{figs3}(f), the gap between them prevents mode coupling. This makes the TDR-NHDR diagram more robust to variations in system size.
This explains why, when $\phi$ deviates from the degenerate point, the edge state transition point finally stabilizes
 at $\theta = \pi/3$ in Fig~\ref{figs7} (a)-(c).

More interestingly, at the degenerate point, the localization of the edge state undergoes a sudden transition from the same end to the opposite ends of the chain, as demonstrated in our analytical solutions and Fig.~\ref{figs3}(b)-(e). 
In contrast, in the non-degenerate case, the distribution of one edge mode gradually shifts from the right to the left, as shown in Fig.~\ref{figs3}(g)-(j). At the edge localization transition point in Fig~\ref{figs3}(h), it displays total delocalized properties. 
This in-gap extended mode is a result of the competition between the non-Hermitian skin effect and band topology, which reveals a promising method for topologically-protected wave control.
It's worth noting that in Wang et al., Nature 608, 50 (2022)~\cite{nature2022morphing}, 
they find a similar phenomenon, which is induced by the competition between a Hermitian-non-Hermitian domain wall and skin effects. They point out that these delocalized states, called "topological morphing modes", are very useful for deforming topological modes into a diversity of shapes.
Our research not only expands and deepens the current understanding of non-Hermitian edge state transitions but also paves the way for new approaches in topological engineering.
}

\subsection{Exact calculation of edge states}\label{exactEdge}

This competition between symmetry and non-Hermitian skin effect can be understood by the analytical solution of the two surface states of our  Hamiltonian in Eq.~(\ref{seqham3}), satisfying
\begin{eqnarray}
 H \sum_{n=1}^{N}\left(a_n|n, a\rangle+b_n|n, b\rangle\right)
 =\epsilon_s \left(a_n|n, a\rangle+b_n|n, b\rangle\right) .
\end{eqnarray}
This gives $2 N$ equations for the amplitudes $a_{n}$ and $b_{n}$ as follows,
\begin{equation}
    \begin{aligned}
        \label{s35}
       \mathrm{for~} n=1,2...,N-1:
       \quad 
       \lambda_2 \text{cos}\theta~a_n
       +(\omega_0+\omega_b \text{cos}(\phi+\Delta\phi)) ~b_n
       +j ~a_{n+1}
       &=\epsilon_s ~b_n;
       \\
        j ~b_n
        +(\omega_0+\omega_a \text{cos}\phi_a) ~a_{n+1}
        +\lambda_1 \text{cos}\theta ~b_{n+1}
        &=\epsilon_s ~a_{n+1};
       \nonumber \\
       \text{for boundaries}:
       \quad\quad\quad\quad
        (\omega_0+\omega_a \text{cos}\phi_a) ~a_1+ \lambda_1 \text{cos}\theta ~b_1
        &=\epsilon_s ~a_1;
        \\ 
        \lambda_2 \text{cos}\theta ~a_N+(\omega_0+\omega_b \text{cos}(\phi+\Delta\phi)) ~b_N
        &=\epsilon_s ~b_N. 
    \end{aligned}
\end{equation}
For the sake of computational convenience, we can decompose the on-site terms $\omega_a \text{cos} \phi_a$ and $\omega_b \text{cos} (\phi+\Delta\phi)$ into a common central frequency and a detuning frequency. This decomposition takes the form: $\omega_a \text{cos} \phi_a=\omega_c+\delta\omega$, $\omega_b \text{cos} \phi_a=\omega_c-\delta\omega$, where $\omega_c=\left(2 \omega_0+\omega_a \text{cos}\phi_a+\omega_b \text{cos}(\phi+\Delta\phi)\right)/2$ and $\delta \omega=\left(\omega_a \text{cos}\phi_a-\omega_b \text{cos}(\phi+\Delta\phi)\right)/2$.
The common part acts as a frequency shift, which does not influence the mode profiles. 
Without loss of generality, we can remove it by switching to a frame rotation at frequency $\omega_c$.  Thus, the Eq.~(\ref{s35}) can be rewritten in:
\begin{equation}
    \begin{aligned}
        \label{s35v2}
       \mathrm{for~} n=1,2...,N-1:
       \quad 
       \lambda_2 \text{cos}\theta ~a_n
       -\delta \omega ~b_n
       +j ~a_{n+1}
       &=\epsilon_s ~b_n;
       \\
        j ~b_n
        +\delta \omega ~a_{n+1}
        +\lambda_1 \text{cos}\theta ~b_{n+1}
        &=\epsilon_s ~a_{n+1};
       \nonumber \\
       \text{for boundaries}:
       \quad\quad\quad\quad
        \delta \omega ~a_1+ \lambda_1 \text{cos}\theta ~b_1
        &=\epsilon_s ~a_1;
        \\ 
        \lambda_2 \text{cos}\theta ~a_N-\delta \omega ~b_N
        &=\epsilon_s ~b_N. 
    \end{aligned}
\end{equation}
where the surface states satisfy $\epsilon_s=\pm \delta\omega$.
When $\delta \omega=0$, we have two zero energy surface states.
The solution to the above equation is
\begin{eqnarray}
\label{seqa5}
a_{n} &=& a_1 \left(\frac{-\lambda_2 \text{cos}\theta}{j}\right)^{n-1}, 
\quad {\rm for\ } n =1, \ldots, N;
\quad\quad 
\nonumber \\
b_{n} &=& b_N \left(\frac{-\lambda_1 \text{cos}\theta}{j}\right)^{N-n} , 
\quad {\rm for\ } n =1, \ldots, N.
\end{eqnarray}
where $a_N={\sqrt{\lambda_2^2 \text{cos}^2\theta-j^2}}/{j}$ and 
$b_1={\sqrt{j^2-\lambda_1^2 \text{cos}^2\theta}}/{j}$ are normalized parameters.
In the thermodynamic limit ($N \rightarrow \infty$), to satisfy the boundary condition $a_N=b_1= 0$, the intercell hopping must be stronger than the intracell hopping, i.e., $\lambda_{1(2)}\text{cos}\theta<j$. 
If $\lambda_{1}\text{cos}\theta<j$ and $\lambda_{2}\text{cos}\theta<j$, we have two zero-energy solutions, which localize at the different boundaries:
\begin{eqnarray}
 |L\rangle=\sum_{n=1}^{N} a_{n}|n, a\rangle, \quad|R\rangle=\sum_{n=1}^{N} b_{n}|n, b\rangle.
\end{eqnarray}
In this region, dominated by band topology, the edge states separately distribute at two ends.

In contrast, if $|\lambda_1 \text{cos}\theta|<|j|$ and $|\lambda_2 \text{cos}\theta|>|j|$ (assuming $|\lambda_2|>|\lambda_1|$), the boundary condition $a_N=0$ requires $a_n=0$ for all $n$, thus the edge states will only takes this form:
\begin{eqnarray}
\quad|R\rangle=\sum_{n=1}^{N} b_{n}|n, b\rangle.
\end{eqnarray}
This result indicates the two edge states will localize at the same edge, which can be understood as driven by the non-Hermitian skin effect.
The interplay and competition between band topology and non-Hermiticity leads to a transition from  non-Hermitian dominance to symmetry dominance in the behavior of the localized boundary modes.
The transition point where the edge states distribute at different ends to the same end is determined by the condition:
\begin{equation}
   \rm{max}\{|\lambda_1 \text{cos}\theta|,
   |\lambda_2 \text{cos}\theta|
   \}=|j|.
\end{equation}

%%%%%%%%%%%%%%%%	

% \bibliography{ref}

%--------------------------------------------------------------------------------

\end{widetext}

\end{document}